\documentstyle[12pt]{article}
\begin{document}

\title{Charge dependence of $NN\rightarrow d\pi$}
\author{J.A. Niskanen and M. Vestama\\
Department of Physics, P.O. Box 9 \\ FIN-00014 University of
Helsinki, Finland}
\maketitle

\begin{abstract}
We calculate the isospin symmetry violating effects to the 
reactions $pp\rightarrow d\pi ^+$ and  $np\rightarrow d\pi ^0$
arising from the different hadron masses and from the Coulomb 
interaction between the positive pion and the deuteron. These
effects are large enough in the cross section and analyzing
power $A_y$ that they should be taken into
account in comparisons of accurate experiments in different
charge channels.
\end{abstract}



\vspace{0.5cm}

Isospin symmetry has been used with great success to relate the
experimentally available pion production reactions
\begin{equation}
pp\rightarrow d\pi ^+
\end{equation}
 and
\begin{equation}
np\rightarrow d\pi ^0.
\end{equation}
 Another charge channel is $nn\rightarrow d\pi^-$, but this is 
impracticable. Further, positive (or negative) pion absorption 
on the deuteron
\begin{equation}
\pi^+ d\rightarrow pp
\end{equation}
can be related to (1) by detailed balance.
 Because the isospin one component of the $np$
initial state has the probability $\frac{1}{2}$, 
then according to this symmetry the cross section for the 
reaction (2) is simply one half of the
$pp$ initiated reaction (1). Relative angular distributions
and spin observables of both reactions should be the same.

However, the isospin symmetry is not exact. Since the masses of the
initial state nucleons as well as the masses of the
charged and neutral pions are different, 
the $T_z=0$ reaction (2) has a 5.9 MeV
lower threshold (CMS) than the $T_z=1$ reaction (1). This has a
potential to be a significant effect as compared with errors of 
modern accurate experiments. Some of this difference can be 
accounted for by different phase space factors in the cross 
sections. In relative observables these factors cancel.

After the removal of the trivial phase space factors, nevertheless,
there remain some charge dependent effects in the transition 
amplitudes themselves. These have three basic origins:

1) Kinematic difference: The relation of the initial momentum to the
final momentum is not exactly the same for different reactions. Even
for the same final momentum (the more relevant in threshold reactions with
a large negative Q-value), the initial momentum in the baryon wave
function is slightly different. The latter enters the transition 
matrix via the $NN$ wave function in a way dependent on the partial 
wave and can, in principle, influence also the relative
observables.

2) The Coulomb interaction enters the channel (1) in both the
initial and the final state, while it is absent in (2).
Presumably for the high energy nucleons it can be neglected,
but it can cause a significant correction in the $d\pi ^+$ final state
in the threshold region. One may safely neglect the magnetic
interaction (although it may be relatively important in the small {\it 
charge symmetry breaking} in $np\rightarrow d\pi ^0$).

3) Charge dependent strong force: A more interesting possibility is charge
dependence in dynamics, i.e. the interaction involved. The lighter and
longer-ranged $\pi ^0$ exchange enters differently in the $pp$ and
$np$ initial states. This gives rise to an isotensor force both in 
$NN$ scattering \cite{henley} and in the $NN\rightarrow \Delta N$ 
transition potential \cite{break} as well as in pion $s$-wave
rescattering. Another source for a charge dependent dynamic 
force can be meson mixings $\pi^0 \leftrightarrow \eta$ and
$\rho^0 \leftrightarrow \omega$. However, the latter violate also
charge symmetry and are presumably smaller.

The aim of this Letter is to study the first two contributions to
charge dependence in pion production. The motivation is an urgent 
need for an estimate because of experimental reasons. The accuracy of
experiments on $np\rightarrow d\pi^0$ is aiming at a level where even
the possible variation from the above sources may be significant in 
comparing different charge channels \cite{lacker}. Precision
measurements of $np$ scattering at intermediate energies often use
pion production and isospin symmetry in normalizing the 
luminosity and the above differences could be a source of a 
systematic error. This in turn
may have implications e.g. to the determination of the (charged)
pion coupling to nucleons, a topic of much controversy recently
\cite{ericson}. Also there is progress in threshold
experiments, where the Coulomb corrections are most relevant 
\cite{bondar,meyer,hutch,cosy}. In other reactions, such as 
$pp\rightarrow pp\pi ^0$ its inclusion has been essential \cite{pppi}. 
Therefore it is
necessary, if not interesting, to get a quantitative estimate of charge
dependence in the $NN\rightarrow d\pi$ reactions. 
Although these violations of the isospin symmetry may be
small, to our knowledge there have not been any systematic calculations
of these to really raise comparisons of experiments above  {\it ad hoc}
reliance on charge independence.

We shall discuss the charge dependent effects in two parts, showing
the change in the observables most likely to be compared (the
differential cross section and the analyzing power $A_y$). First we
calculate the kinematic effect due to the change of the initial state
making all comparisons for the same final pion momentum in the CMS. 
Then we proceed to
incorporate the final state Coulomb corrections in  different
approximate ways. The first one is simply to apply the Coulomb 
penetration factors $| C_l|^2$, and in the second one we replace 
the plane wave pion in the matrix elements by Coulomb scattering 
wave functions. Finally, the effect of the finite charge distribution
 is also estimated.
In principle, one should consider also the polarizability of the
deuteron. However, we end up with the conclusion that this is most
likely a small effect, much smaller than the already small ones that
we include. Technically including the polarizability would be also much
harder.

A detailed discussion of the model used can be found in 
Ref. \cite{ppdpi}. In addition to the direct 
production from the initial $NN$ states, we include
 by the coupled-channels method $p$-wave pion rescattering 
through the $\Delta (1232)$ isobar (dominant above 350 MeV in
$p$-wave production from $^1D_2$) and also $s$-wave 
rescattering (dominant at threshold). The Reid soft core wave
function is used for the deuteron. This model gives rather
good predictions to the cross sections and spin observables
through the $\Delta$ region, but overestimates threshold
production by about 50 \% \cite{comment}. 

 In the naive 
quark model the mass difference between the $NN$ channels
and the $\Delta N$ channels does not depend on the charge
\cite{break}, because the same six quarks are involved in both. 
This means that the kinetic energy required for the excitation 
of a $\Delta N$ intermediate state is the same for the $np$ 
and $pp$ initial states. Therefore no difference in the 
transition $NN\rightarrow \Delta N$ arises from the baryon 
masses at the same
incident nucleon kinetic energy. It was also pointed out
in Ref. \cite{break} that in lowest order the internal Coulomb effects
should have a similar effect to the total masses in all channels.

The production cross section of $pp\rightarrow d\pi^+$ 
in the centre-of-mass system can be expressed in terms of 
the matrix elements of the production operator as
\begin{equation}
\frac{d\sigma_{\rm prod}}{d\Omega}=\frac{1}{(2\pi)^2}\frac{\omega_q
E_dE_1E_2q}{ps}\frac{1}{4}\times\sum_{\mu SM}\mid<\psi_d^\mu\mid
H^\pi\mid\phi^{SM}>\mid^2 = P_{pp}\times R_{pp}.
\end{equation}
Here $q$ is the pion momentum and $p$ the initial nucleon 
momentum. The energies $E_1$ and $E_2$ are the initial nucleon  
energies, while $\omega_q$ and $E_d$ are the final state energies
for the pion and deuteron, 
and $s$ is the Mandelstam variable. All kinematic quantities 
refer to the centre-of-mass system. The cross section is
explicitly factorized to the phase space factor $P_{pp}$ and
the sum of squared matrix elements $R_{pp}$. For the $np$ 
reaction (2) one needs also the additional isospin factor 
$\frac{1}{2}$ relative to the reaction (1) and
our factorization is $2\sigma(np)=P_{np}\cdot R_{np}$,
so that also $R_{np}$ corresponds to similar pure isospin 
one matrix elements as $R_{pp}$ does.
The notation is valid for both the differential and integrated
cross sections. In the former the $R$ is just angle dependent.

The absorption cross section (3)
\begin{equation}
\frac{d\sigma_{\rm abs}}{d\Omega}=\frac{1}{(2\pi)^2}\frac{\omega_q
E_dE_1E_2p}{qs}\frac{1}{3}\times\sum_{\mu SM}\mid<\psi_d^\mu\mid 
H^\pi\mid\phi^{SM}>\mid^2,
\end{equation}
is related to its inverse by detailed balance, i.e. by the
factor $\frac{4}{3}p^2/q^2$. Here
the integrated absorption cross section needs an additional 
factor $\frac{1}{2}$ from the identity of the protons
in the case of positive pion absorption. The $np$ reactions 
need the same factor $\frac{1}{2}$ for the isospin
already in the differential cross section, so the two total 
absorption cross sections should be equal for exact isospin
symmetry.
A similar factorization as above is possible also here with
the same quantities $R$, but different phase space factors.
Of course, the inverse of (2) is hardly experimentally
feasible, but we keep it for symmetry. Often it is better
to present the cross sections in terms of $\sigma_{\rm abs}$
and Fig. 1 facilitates a comparison in this presentation.

Both the phase space factor ($P$) and the sum of the squared 
matrix elements ($R$) depend on the charge via masses and
different momenta. One sees easily the following relation 
for the difference of the production cross sections (both 
differential and total)
\begin{equation}
 \delta\sigma \equiv 2\sigma(np)-\sigma(pp)
=\frac{P_{np}+P_{pp}}{2}(R_{np}-R_{pp})
+(P_{np}-P_{pp})\frac{R_{np}+R_{pp}}{2}.
\end{equation}
A similar equality holds for the differential absorption 
cross sections with just different phase space factors $P$. 
(For the total absorption cross section
difference one should use 
$\sigma_{\rm abs}(np)- \sigma_{\rm abs}(pp)$.)
Also the sum of the cross sections can be expressed 
exactly as
\begin{equation}
 2\sigma_{\rm av}\equiv 2\sigma(np)+\sigma(pp)
=\frac{P_{np}+P_{pp}}{2}(R_{np}+R_{pp})
+(P_{np}-P_{pp})\frac{R_{np}-R_{pp}}{2},
\end{equation}
where the latter term is only a high order correction. 
Neglecting it as a good approximation one gets finally
the relative change of the cross section simply as the 
sum of the relative changes of the phase space factor and 
of the squared matrix elements
\begin{equation}
\frac{\delta\sigma}{\sigma_{\rm av}}\approx
\frac{\delta R}{R_{\rm av}} + \frac{\delta P}{P_{\rm av}}
\label{ds}
\end{equation}
with $P_{\rm av}$ and $R_{\rm av}$ average quantities between
the two reactions. 

From the above discussion it seems most reasonable
to present the theoretical results in terms of the 
calculated ratio $\delta R /R_{\rm av}$ with
\begin{equation}
 \delta R=
R_{np}-R_{pp}=\sum_{\mu SM}2\mid<\psi_d^\mu\mid H^\pi\mid\phi^{SM}>_{np}
\mid^2-\mid<\psi_d^\mu\mid H^\pi\mid\phi^{SM}>_{pp}\mid^2,
\end{equation}
which can be used above to estimate the relative
difference in the cross sections. 
Replacing the average $\sigma_{\rm av}$ by either 
$\sigma_{pp}$ or $\sigma_{np}$ one can then get an estimate 
of the other cross section with a very small error. 
This prescription
presumably  minimizes the systematic effects of possible 
normalization deviations of the model calculation from 
experiments, since if the cross section is overestimated,
likely also the difference is. This expectation will be borne 
out later in the discussion of the model dependence. 
Also the choice of the placement of the constant statistical
factors either in $P$ or $R$ now actually becomes irrelevant 
in these relative quantities.

Fig. 1 shows $\delta R/R_{\rm av}$ and  $\delta P/P_{\rm av}$ 
in different cases as a function of $\eta=q/m_{\pi^+}$. First, 
the dashed curve shows only the mass difference 
effect. At threshold with delicate cancellations of the
oscillatory $NN$ wave function, the long-ranged
Galilean invariance term 
$\propto (\vec p + \vec p')\cdot (\vec\sigma_1 
+ \vec\sigma_2)$ \cite{ppdpi} yields a larger amplitude for
the reaction (2). Although this is only a quarter of the
dominant $s$-wave rescattering amplitude, this source is 
still responsible for about $+2$ percent units in $\delta R/
R_{\rm av}$. Further, the $pp$ momentum for the same pion
momentum close to threshold is about two percent larger than
the $np$ momentum. Because of the wave function normalization
as $\sim j_L(pr + \delta_L)$ this makes the cross section
of the reaction (1) 4 \% smaller than (2). The wide and deep
dip in Fig. 1 is due to the $\Delta$ isobar, which  increases 
both cross sections. However, for any given pion momentum 
the necessary $pp$ kinetic energy is higher 
and, therefore, closer to the resonance as long as the energy 
is still below the $\Delta$ threshold. (At threshold the 
difference in the distance to the $\Delta$ is about 5 \%.)
Therefore, the reaction (1) becomes stronger than (2) 
just below the $\Delta$ region, while the effect in 
$\delta\sigma$ is reversed above it, where the $pp$ energy is
further from the $\Delta$. Here it is essential to remember
that the distance of the $\Delta N$ mass from the two-nucleon
mass is the same for both $np$ and $pp$ channels as discussed
above before Eq. (4). This strong energy dependence
of the cross section difference may be somewhat unexpected
but perfectly understandable. In size  the mass effect 
alone is, in fact, already comparable to today's best 
experimental uncertainties. 
 
However, in $pp\rightarrow d\pi^+$ there is also the Coulomb force 
present. Although for the high-energy initial state this can   
arguably be neglected, it 
is important in the final state close to threshold.
It may be noted that for larger pion energies both Coulomb effects 
become comparable, since the $pp$ reduced mass is much larger
that the pion mass. However, the Coulomb effect in the initial 
$pp$ state is always smooth and rather small
and we do not consider it here. The two become about equal in the
neighbourhood of $\eta\approx 1.2$. 

In this work we consider the Coulomb effect in three different ways. 
First we take it into account in different pion partial waves
 by the Coulomb penetration factors  \cite{abra}
\begin{equation}
C_l(\xi) = \left[\prod_{n=1}^{l}(1+\frac{\xi^2}{n^2})
\right]^{\frac{1}{2}}\, {C_0(\xi)}\, .
\end{equation}
Here the denominator $(2l+1)!!$ present also in the series expansion 
of the spherical Bessel functions has been removed for a direct
comparison with plane waves.  As can be seen in e.g. the Appendix of the 
first of Refs. \cite{ppdpi}, for plane wave pions some spherical Bessel 
functions $j_l(qr/2)$ enter the overlap integrals, where $l$ is not 
necessarily the same as the pion angular momentum $l_\pi$ relative to 
the deuteron. We multiply these by the above $C_l(\xi)$. As 
 $\xi=\alpha m_{\rm red}c^2/(\hbar cq)\approx 0.0068/\eta$ 
is rather small for any presently reported value of $\eta$, 
all penetration factors are 
practically the same as $C_0(\xi)=[2\pi\xi(e^{2\pi\xi}-1)^{-1}]^{1/2}$ 
($m_{\rm red}$ is the $\pi d$ reduced mass). 
The difference with the inclusion also of this 
effect is shown by the solid curve in Fig. 1. Below the $\Delta$ 
resonance the negative difference is partly cancelled off 
because of the Coulomb 
suppression of the $d\pi^+$ final state, while at threshold 
$\delta R /R_{\rm av}$ approaches 2.

Also for completeness, the dotted and dash-dot curves show the
relative change of the phase space factors $\delta P /P_{\rm av}$ 
for the production and absorption reactions, respectively. The 
relative cross section difference $\delta\sigma /\sigma_{\rm av}$ 
can now be obtained simply by adding to $\delta R/R_{\rm av}$ 
the corresponding $\delta P/P_{\rm av}$, as shown in Eq. \ref{ds}.

A relatively standard way of including Coulomb corrections has been
to apply to the cross section the penetration factor $C_0^2$.
Using only this gave results indistinguishable
from the procedure described above.
Also the use of $C_{l_\pi}$ in various pion amplitudes directly did not
deviate significantly. Another way (though still not exact)
of including the Coulomb effect is to explicitly distort the 
Bessel functions to the Coulomb functions $j_l\rightarrow F_l$. However, 
one should note that this has some difficulty as having different cases: 
either an emission of a charged pion with also a charged nucleon (as 
$\Delta^{++}\rightarrow p\pi^+$) or a proton or a $\Delta^+$
producing a $\pi^+$, which 
interacts by the Coulomb force (possibly combined with s-wave 
rescattering) with the {\em second} nucleon. A reasonable estimate may 
be obtained by a simple replacement of the wave functions which should 
maximize the effect of this change. Only in case of serious disagreement 
with the earlier result is there reason for worry. However, the results
obtained in this way were essentially the same as those with the simpler
multiplicative penetration factors.

A third way of looking at the Coulomb effects is to consider also 
the finite size of the charge distribution of the deuteron. For this 
purpose we have modified the Coulomb potential due to the charge 
distribution of the Reid deuteron wave function to
\begin{equation}
 V_{\rm Coul}^m(\vec r_\pi) = e^2\int d\Omega\, dr\,
\frac{u^2(r) + v^2(r)} {|\vec r_\pi - \vec r/2|}
\end{equation}
and used it in a Schr\"odinger equation to numerically solve 
the relevant pion wave functions to be used in the matrix elements. 
This approach neglects the nonspherical parts of the potential
due to the $D$ state, which would cause a coupling between
different pion partial waves. The effect of charge extension 
is to very slightly increase 
the penetration (i.e. weaken the repulsion), but contrary to the 
expectation of Ref. \cite{reitan} the change is not large enough 
at any energy to warrant an additional calculation of the 
observables (at the level 1.5\% for $s$ waves and 0.5\% for
$p$ waves). Table I shows the penetration factors $C_0^2$ 
and $C_1^2$ for a point-like and extended charge at a 
few low momenta. In the threshold parametrization $\sigma_{\rm prod}
=\alpha\eta +\beta\eta^3$ these have been used in the past
to scale the parameters $\alpha$ and $\beta$.
 Even though the deuteron is rather extended
by nuclear scale, its spatial
 variation is much faster than that of the
pion wave function at very low energies. So basically threshold
pions then feel a point charge. Further above threshold the
Coulomb force loses importance. So either way the influence 
of the charge distribution is rather negligible to the reaction.
This insensitivity to the details of the charge distribution 
also suggests that the effect of deuteron polarizability
can be rather safely neglected.  It may be noted that the present
finite-size effects are only about a quarter of the results of
Ref. \cite{meyer}, which, integrating only to $r_\pi$, missed 
part of the charge
\cite{heimberg}.

Fig. 2 shows the corresponding changes in the analyzing power $A_y$
at 90$^\circ$ for a polarized beam, 
which is rather representative, since its angular shape 
changes slowly and systematically. The effect of the mass difference
alone is rather small, as can be seen from the dashed curve. The
result remains indistinguishable from this, if also the Coulomb 
penetration factors are applied on the pion wave functions. This
means, of course, that the effects are similar in different 
partial waves and do not change relative phases. However, adding 
the asymptotic Coulomb phase $\sigma_{l_\pi}$ to the amplitudes 
does change the relative phases and has a significant effect
below $\eta\approx 0.8$ as can be seen  from the solid curve. To
emphasize the origin of this effect the dotted curve shows still
the difference $A_y(\sigma_{l_\pi}=0) - A_y(\sigma_{l_\pi}\neq 0)$ 
for the pure $pp$ case; clearly the Coulomb phases
$\sigma_{l_\pi}$ dominate the change of $A_y$. 

It was argued above that the chosen presentation of the results as the
ratio $\delta R/R_{\rm ave}$ would minimize the role of theoretical
uncertainties and model dependence. We tested this in various ways.
Using the Paris deuteron wave function instead of the Reid soft core
one, threshold production ratio changes by about +0.6\% and $\delta
A_y$ by +0.01. In the Paris wave function there is a significantly 
softer hard core. At higher energies the changes are smaller.

Next the initial state interaction was changed strongly by
enhancing the $NN\rightarrow N\Delta$ transition
potential by 20\%. In the pion production energy region
 this changes the $NN$ phase shifts 
by several degrees and the cross section is changed by 10-25\%. 
 Nevertheless, at threshold the effects in both $\delta R/R_{\rm ave}$ 
 and $\delta A_y$ were totally negligible.  
 In the $\Delta$ region the change in the ratio
$\delta R/R_{\rm ave}$  was about -1\% and in $\delta A_y$
-0.005. Such a large variation of the
transition potential is, in fact, not allowed. Namely,
in Ref. \cite{trans} a strong point was made that the $\Delta$
strength can be well fixed by the total pion production cross section
in the $\Delta$ region, so there is not much uncertainty from this
source. As an ultimate test the $\Delta$ was switched off and only the
standard
Reid soft core $NN$ potential used. At threshold this has an effect 
of reducing the cross section by a factor of $\frac{1}{3}$
\cite{comment}. Even with this massive overall change
 the threshold value of the $\delta R/R_{\rm ave}$ 
increased by only 2 percent units to 9\%, and then this
quantity decreases monotonously from this value with increasing
energy being about 3\% at $\eta = 1.8$. This gives a measure of 
the relative overall importance of the $\Delta$ at threshold.
At higher energies its role is more pronounced, as testified 
by the strong dip, which exclusively arises from approaching and 
passing the resonance. Here the $\Delta$ effect is about as
striking as in the total cross section itself. However, as discussed
above, there is little uncertainty in the role of the $\Delta$.
On the other hand, alterations of only the nucleon correlations 
have much less effect as was seen e.g. in Ref. \cite{trans}.

Recently arguments have arisen about the importance of off-shell
rescattering of the pion. Presently we consider only a monopole form
factor in the whole rescattering process with the $\pi N$ 
scattering amplitude taken from the on-shell analysis of Ref.
\cite{kh80}, which assumes charge independence. While the form
factor should be reasonable for the isovector
scattering presumably mediated by $\rho^\pm$ mesons, it may not 
be sufficient for the isospin symmetric isoscalar channel. Varying
the cut-off mass by a few hundred MeV had very little effect
on the difference between the two reactions, while the cross section
itself changed by tens of percent at threshold. However, 
at a deeper level the different masses of the intermediate 
mesons could have some charge dependent effect in the off-shell 
scattering amplitude, but such a calculation would require an
actual model of $\pi N$ scattering, based perhaps on chiral 
perturbation theory. One would expect some effect also from the 
different pion masses in the intermediate pion propagator. 
However, both of these effects belong to the class of dynamic 
isotensor forces outside the scope of this Letter. Work on such
interactions is in progress. 
 
In summary, we have studied the changes in the total cross 
sections and analyzing powers $A_y(90^\circ)$ of the reactions
$pp \rightarrow d\pi^+$ and $np \rightarrow d\pi^0$ caused by
the "trivial" differences in the masses and the Coulomb 
interaction. Depending on energy, these can be at a several 
percent level with varying sign, and in detailed comparisons 
of data from different reactions they should be taken into account
as well as in calibrating neutron beams by pion production.
While the Coulomb effect can be treated simply by  
multiplicative factors $C_0(\xi)^2$, the mass difference causes
a nontrivially energy dependent effect, which requires an actual
model calculation, the task accomplished here.
The angular distributions change significantly less than the
above observables setting an overall scale of the changes.
Overall, our estimation of the uncertainties of the present 
calculation (within its scope to consider only
the mass differences in kinematics and the Coulomb effect) would be
about one percent unit for the $\delta R/R_{\rm ave}$ and 0.01 for
$\delta A_y$. 
\\[3mm]

{\em Acknowledgement:} M.V. was partially supported by the
Research Institute for Theoretical Physics, University of
Helsinki. We acknowledge useful communications with J. Blomgren,
H. Lacker and L. Schmitt.

%\newpage

%\vspace{1cm}
\newpage

\centerline{TABLE}
\vspace{3mm}

\begin{table}[h]
\caption{The penetration factors for a point-like charge
and an extended distribution (labelled e). } 
\begin{center}
\begin{tabular}{|c|c|c|c|c|} \hline
$\eta$ & $C_0^2$ & ${C_0^e}^2$ & $C_1^2$ & ${C_1^e}^2$\\ 
\hline 
0.01 & 0.061 & 0.061 & 0.089 & 0.089  \\ 
0.02 & 0.286 & 0.291 & 0.319 & 0.321  \\ 
0.04 & 0.560 & 0.568 & 0.576 & 0.579  \\  
0.06 & 0.686 & 0.696 & 0.695 & 0.698  \\   
0.08 & 0.757 & 0.768 & 0.762 & 0.766  \\  
0.10 & 0.802 & 0.813 & 0.805 & 0.809  \\
0.15 & 0.864 & 0.877 & 0.866 & 0.871  \\
0.20 & 0.897 & 0.910 & 0.898 & 0.903  \\ 
0.30 & 0.931 & 0.943 & 0.931 & 0.936  \\ \hline
\end{tabular}
\end{center}
\end{table}

\vspace{2cm}

\centerline{FIGURE CAPTIONS}
\vspace{3mm}

\noindent Fig. 1: Relative changes in the sum of the squared 
matrix elements ($R$) and phase space factors ($P$). Dashed curve:
the mass difference effect in $R$; solid: the full result for $R$ 
including 
also the Coulomb repulsion; dotted: the change in $P$ for production;
dash-dot: the change in $P$ for absorption. \\

\noindent Fig. 2: Differences between $\vec np\rightarrow d\pi^0$ and
$\vec pp \rightarrow d\pi^+$ in the analyzing power $A_y$ at 90$^\circ$.
Dotted: the mass difference effect alone (indistinguishable if
also the penetration factors are included);  solid:
 the Coulomb phase $\sigma_{l_\pi}$ included; dashed:
the resulting change in the $\vec pp \rightarrow d\pi^+$ reaction,
if the Coulomb phases are switched off. 


\begin{thebibliography}{99}

\bibitem{henley} E. M. Henley and G. A. Miller in Mesons
in Nuclei, Vol. I, eds. M. Rho and D. H. Wilkinson,
(North Holland, Amsterdam, 1979) p. 416;
T. E. O. Ericson and G. A. Miller, Phys. Lett. 132 B (1983) 32;
C. Y. Cheung and R. Machleidt, Phys. Rev. C 34 (1986) 1181.
\bibitem{break}J. A. Niskanen and A. W. Thomas, Phys. Rev. C 
 37 (1988) 1755.
\bibitem{lacker} H. Lacker, private communication.
\bibitem{ericson} T. E. O. Ericson {\it et al.}, Phys. Rev. Lett.
 75 (1995) 1046 and references therein.
\bibitem{bondar} A. Bondar {\it et al.}, Phys. Lett. 
 B 356 (1995) 8.
\bibitem{meyer} P. Heimberg {\it et al.}, Phys. Rev. Lett.
 77 (1996) 1012.
\bibitem{hutch} D. A. Hutcheon {\it et al.}, Nucl. Phys. 
 A 535 (1991) 618; E. Korkmaz {\it et al.}, Nucl. Phys. 
 A 535 (1991) 637.
\bibitem{cosy} M. Drochner {\it et al.}, Phys. Rev. Lett.  77
(1996) 454.
\bibitem{pppi} H. O. Meyer {\it et al.}, Nucl. Phys. A 539 (1992)
633; G. A. Miller and P. U. Sauer, Phys. Rev. C 44 (1991) 1725;
J. A. Niskanen, Phys. Lett.  B 289 (1992) 227.
\bibitem{ppdpi}J. A. Niskanen, Nucl. Phys.  A 298 (1978) 417,
  Phys. Lett. 141 B (1984) 301.
\bibitem{comment} J. A. Niskanen, Phys. Rev. C  53 (1996) 526.
\bibitem{abra} M. Abramowitz and I. Stegun, Handbook of 
Mathematical Functions,  3rd printing, (Natl.  Bureau of 
Standards, 1966).
\bibitem{reitan} A. Reitan, Nucl. Phys.  B 11 (1969)170.
\bibitem{heimberg} P. Heimberg, private communication.
\bibitem{trans} J. A. Niskanen and P. Wilhelm, Phys. Lett.
 B 359 (1995) 295.
\bibitem{kh80} R. Koch and E. Pietarinen, Nucl. Phys. A 336 (1980)331.
\end{thebibliography}
\end{document}